\documentstyle[preprint,aps,epsbox]{revtex}

\newcommand{\beq}{\begin{equation}}
\newcommand{\eeq}{\end{equation}}
\newcommand{\bea}{\begin{eqnarray}}
\newcommand{\eea}{\end{eqnarray}}
\newcommand{\nn}{\nonumber\\}
\newcommand{\mpl}{{m_{\rm Pl}}}
\newcommand{\mpll}{m_{\rm Pl}^{~2}}
\newcommand{\xx}{1-6\xi}
\newcommand{\gsim}{\mbox{\raisebox{-1.0ex}{$\stackrel{\textstyle >}{\textstyle \sim}$}}}
\newcommand{\lsim}{\mbox{\raisebox{-1.0ex}{$\stackrel{\textstyle <}{\textstyle \sim}$}}}

\begin{document}

\draft
\tighten
\preprint{YITP-98-80, hep-ph/9901336}
\title{Topological inflation\\ 
induced by a non-minimally coupled massive scalar field}
\author{Nobuyuki Sakai\thanks{
E-mail: sakai@yukawa.kyoto-u.ac.jp; Fax: +81-75-753-7010}
and Jun'ichi Yokoyama}
\address{Yukawa Institute for Theoretical Physics, Kyoto University, Kyoto
606-8502, Japan}
\date{revised 12 April 1999}
\maketitle

\begin{abstract}
\noindent
We reanalyze cosmology of a non-minimally coupled massive scalar field
which was originally discussed by Futamase and Maeda in the context 
of chaotic inflation scenario. We find a new type of 
inflationary solution where inflation occurs inside a domain 
wall. This new solution relaxes constraints on the coupling constant 
for successful inflation.

\vskip1cm \noindent
PACS number(s): 98.80.Cq, 04.50.+h 
\end{abstract}

\newpage

Among various inflationary models \cite{rev}, Linde's chaotic inflation 
\cite{Lin} is a simple and natural scenario, which is induced by 
a minimally coupled scalar field with a polynomial potential. From a 
viewpoint of quantum field theory in curved spacetime, however, we expect 
that the scalar field 
may couple with spacetime curvature. Futamase and Maeda \cite{FM} investigated 
how the nonminimal coupling term $(1/2)\xi{\cal R}\phi^2$ affects 
realization of chaotic inflation, and derived constraints on the coupling 
constant $\xi$ from the condition for sufficient inflation. For the potential (i) 
$V(\phi)=(1/2)m^2\phi^2$, they obtained $|\xi|\lsim 10^{-3}$, while for 
the model (ii) $V(\phi)=(1/4)\lambda\phi^4$, 
$\xi\lsim 10^{-3}$ \cite{FM}. Here, we have adopted the sign convention
such that the conformal coupling corresponds to $\xi=+1/6$. 

For the model (ii) or double-well potentials which includes a quartic 
term, density perturbations have been calculated by several 
authors independently \cite{phi4}, who have argued that the fine-tuning problem 
of the self-coupling constant $\lambda$ ($\lsim 10^{-13}$) is loosened 
by the non-minimal coupling term 
if $|\xi|$ is large enough. Their results are confirmed by a more 
rigorous calculation by Makino and Sasaki \cite{MS}. Recently, 
constraints from tensor perturbations are also discussed \cite{KF-HN}.

The model (i), on the other hand, has not been highlighted because the 
result by Futamase and Maeda indicates that it needs a kind of 
fine-tuning, $|\xi|\lsim 10^{-3}$, just from the condition for a
sufficient amount of inflation. In this paper, however, we shall show that 
another type of inflation may be possible if $\xi<0$, which relaxes the
constraint on $\xi$. 

The model with a non-minimally coupled massive scalar field is 
described by the action,
\begin{equation}
{\cal S} =\int d^4 x \sqrt{-g} \left[\frac{\mpll}{16\pi}{\cal R}
    -\frac12\xi\phi^2{\cal R}-\frac12(\nabla\phi)^2-V(\phi)\right]
~~ {\rm with} ~~ V(\phi)=\frac12 m^2\phi^2,
\end{equation}
where $\mpl$ is the Planck mass, and we concentrate on the case $\xi<0$.
Following Futamase and Maeda \cite{FM}, we apply the conformal transformation,
\beq
\hat g_{\mu\nu}=(1+\psi)g_{\mu\nu}
~~ {\rm with} ~~ \psi\equiv-\xi\kappa^2\phi^2
~~ {\rm and} ~~ \kappa^2\equiv{8\pi\over\mpll},
\eeq
and introduce a new scalar field $\Phi$ as
\bea\label{Phi}
\Phi&\equiv&\int d\phi{\sqrt{1+(\xx)\psi}\over 1+\psi} \nn
&=&{1\over\kappa}\left[\sqrt{{\xx\over-\xi}}
          {\rm arcsinh}\sqrt{(\xx)\psi}
        +\sqrt{\frac32}\ln\left\{{\sqrt{1+(\xx)\psi}-\sqrt{-6\xi\psi}\over
          \sqrt{1+(\xx)\psi}+\sqrt{-6\xi\psi}}\right\}\right].
\eea
Then the model is described by the Einstein gravity with a canonical 
scalar field:
\begin{equation}\label{hatS}
  \hat{\cal S} = \int d^4 \hat x \sqrt{-\hat g}
    \biggl[\frac{\hat{\cal R}}{2\kappa^2}-{1\over2}(\hat\nabla\Phi)^2
     -\hat V(\Phi) \biggr]
~~ {\rm with} ~~ \hat V(\Phi)={V(\phi)\over(1+\psi)^2}.
\end{equation}
Taking the background as the spatially flat Friedmann-Robertson-Walker
spacetime,
\beq
d\hat s^2=-d\hat t^2+\hat a^2(\hat t)d{\bf x}^2=(1+\psi)[-dt^2+a^2(t)d{\bf x}^2],
\eeq
the field equations for the homogeneous parts read
\begin{equation}\label{Feq1}
\hat{H}^2\equiv\left({\hat{a}_{,\hat t}\over\hat{a}}\right)^2=
{\kappa^2\over3}\left(\frac12\Phi_{,\hat t}^{~2}+\hat V\right),~~
\Phi_{,\hat t\hat t}+3\hat H\Phi_{,\hat t}+\hat V_{,\Phi}=0,
\end{equation}
where $_{,\hat t}\equiv d/d\hat t$ and $\hat V_{,\Phi}\equiv d\hat V/d\Phi$.

Thanks to the standard form of the field equations (\ref{Feq1}), we can discuss 
the qualitative behavior of $\Phi$ (or $\phi$) in terms of the potential 
shape. We depict $\hat V(\Phi)$ for $\xi<0$ in Fig. 1. 
A distinguished feature of this potential is that it has a maximum 
at $\Phi=\Phi_{\rm max}$ corresponding to $\psi=1$ or 
$\phi=1/\kappa\sqrt{-\xi}\equiv\phi_{\rm max} $. Hence, if the initial 
value of the scalar field, $\Phi_i$, is larger than $\Phi_{\rm max}$ and 
if the energy density of the scalar field
$E_{\Phi}=\Phi_{,\hat t}^{~2}/2+\hat V(\Phi)$ is below $\hat V(\Phi_{\rm 
max})$, $\Phi$ cannot reach the origin $\Phi=\phi=0$, but it will run away to 
infinity as long as the universe is expanding in the conformal frame 
with $\hat H>0$. In this case, because the potential has an asymptotic form,
\beq
\hat V={m^2\over2\xi^2\kappa^2}e^{-\sqrt{{2\over\hat p}}\kappa\Phi}
 ~~ {\rm with} ~~
\hat p\equiv{\xx\over-2\xi}>3 ~~ {\rm for} ~~ \Phi\gg\Phi_{\rm max},
\eeq 
we find power-law inflation \cite{LM} in the conformal frame,
\beq
\hat a\propto\hat t^{\hat p},~~
\kappa\Phi=\sqrt{\hat p\over 2}\ln{2m^2\hat t^2\over(\xx)(3-16\xi)}.
\eeq
In the original frame, on the other hand, the corresponding asymptotic 
solution is 
exponential inflation,
\beq\label{DS}
a\propto e^{H_{as}t},~~ \phi\propto e^{\alpha mt}
~~ {\rm with} ~~ 
H_{as}\equiv{(1-4\xi)m\over\sqrt{-2\xi(\xx)(3-16\xi)}},~~
\alpha\equiv{\sqrt{-2\xi\over(\xx)(3-16\xi)}}.
\eeq
This inflating region does not enter a reheating phase, and hence it 
cannot lead to the present universe. Nevertheless, this solution 
arouses our interest because the spacetime approaches exact de Sitter while 
the scalar field also increases exponentially. To check this result, 
one may also examine the field equations in the original frame \cite{MS},
\beq\label{Feq2}
H^2(1+\psi)={\kappa^2\over3}
  \left(\frac12\phi_{,t}^{2}+V+6\xi H\phi\phi_{,t}\right),~~
\phi_{,tt}+3H\phi_{,t}+6\xi(H_{,t}+2H^2)+V_{,\phi}=0.
\eeq
One can easily confirm that the solution (\ref{DS}) satisfies the 
above equations in the asymptotic regime where $1+\psi$ in the
right-hand-side (RHS) 
of the first equation can be approximated by $\psi$.

In any case, the above runaway solution is irrelevant to our Universe, 
and Futamase and Maeda \cite{FM} obtained a constraint $|\xi|<10^{-3}$ 
so that $\phi_{\rm max}>5\mpl$ and the initial value of $\phi$ 
required for sufficient chaotic inflation $\phi_i\sim 5\mpl$ lies on 
the left of the potential peak.

We can argue, however, that sufficient inflation may be possible even if 
$\phi_{\rm max}<5\mpl$ because the plateau around $\phi=\phi_{\rm max}$ may
cause another channel of inflation. Seen in the conformal frame, some 
domains of the universe relax to $\Phi=0$ and others run away to 
$\Phi\rightarrow\infty$ as the universe expands. In between these two 
classes of regions exist domain walls where large potential energy 
density $\hat V(\Phi_{\rm max})$ is stored. If the curvature of 
the potential is sufficiently small there, such domain walls will 
inflate. This is nothing but topological 
inflation proposed by Linde and Vilenkin \cite{L-V}. A similar phenomenon has been found
in the model with higher-curvature gravity \cite{EKOY}: a potential maximum 
appears in the conformal frame which permits topological inflation.

Let us discuss whether inflation can take place at $\phi=\phi_{\rm max}$, 
following the arguments of Linde and Vilenkin \cite{L-V}, which have 
been verified by numerical analysis \cite{Sak}.
The conditions for slow-roll inflation 
($|\Phi_{,\hat t\hat t}|\ll|3\hat H\Phi_{,\hat t}|$ and
$\Phi_{,\hat t}^{~2}/2\ll\hat V$) are equivalent to
\begin{equation}\label{slow1}
\left|{\hat V_{,\Phi}\over\kappa\hat V}\right|\ll\sqrt{6}, ~~
\left|{\hat V_{,\Phi\Phi}\over\kappa^2\hat V}\right|\ll3.
\end{equation}
For the potential in (\ref{hatS}), inequalities (\ref{slow1}) lead to
\begin{equation}\label{slow2}
{|1-\psi|\over\sqrt{1+(\xx)\psi}\kappa\phi}\ll\sqrt{\frac32}, ~~
{|1-6\psi-(5-35\xi)\psi^2+2(\xx)\psi^3|\over[1+(\xx)\psi]^2\kappa^2\phi^2}
\ll\frac32,
\end{equation}
where we have used (\ref{Phi}). At the potential maximum, the first condition 
is trivially satisfied; hence we have only to check the second 
condition. Substituting $\phi=\phi_{\rm max} ~ (\psi=1)$ into the second 
one in (\ref{slow2}), we find
\beq\label{cond1}
{|\xi|\over1+3|\xi|}\ll\frac 34.
\eeq
The similar relation is derived from the condition that the thickness of the 
wall characterized by the curvature scale of the potential at the 
maximum, $\hat R_w$, is greater than the horizon, $\hat H^{-1}$ \cite{L-V,EKOY}:
\beq\label{cond2}
\hat R_w\hat H=
\sqrt{\kappa^2\hat V(\Phi_{\rm max})\over 3\hat V_{,\Phi\Phi}(\Phi_{\rm max})}
=\sqrt{{1+3|\xi|\over 12|\xi|}}\gsim 1.
\eeq
Inequalities (\ref{cond1}) and (\ref{cond2}) suggest that inflation actually 
takes place at the top of the potential if $|\xi|\ll 1$. Once inflation 
sets in, it continues forever inside a domain wall. This solution indicates
exponential inflation both in the conformal frame and in the original 
frame, because the conformal factor changes only slowly during slow 
roll-over topological inflation.

In order to obtain more precise conditions for sufficient inflation, we 
solve field equations (\ref{Feq1}) numerically. We assume initial 
values as $\Phi_i=\delta\Phi_{\rm Q}\equiv\hat H/2\pi$ and 
$\dot\Phi_i=0$, and observe the e-fold number of inflation after the classical 
dynamics dominates over quantum fluctuations, {\it i.e.}, 
$|\dot\Phi|/\hat H>\delta\Phi_{\rm Q}$. Sufficient expansion requires typically
\beq
{a_f\over a_c}=\sqrt{{1+\psi_c\over1+\psi_f}}{\hat a_f\over\hat a_c}
>e^{65}, 
\eeq
where $\hat a_c$ denotes the scale factor in the conformal frame when
$|\dot\Phi|/\hat H=\delta\Phi_{\rm Q}$,
and $\hat a_f$ that when the slow-roll conditions break down. The allowed 
region is plotted in Fig. 2. This shows that inflation continues 
sufficiently even if $|\xi|\cong0.1$, contrary to the previous result \cite{FM}.

Next, we investigate density perturbations generated during 
inflation, and constrain the model from the 4yr COBE-DMR data 
\cite{Ben}. Because Makino and Sasaki \cite{MS} showed that the density 
perturbation in the original frame exactly coincide with that in the 
conformal frame, we can easily calculate the amplitude and the 
spectral index with the well-known formulas. The amplitude of 
perturbation on comoving scale $l=2\pi/k$ is given in terms of Bardeen's 
variable $\Phi_A$ \cite{Bar} as
\begin{equation}
\Phi_A\left(l={2\pi\over k}\right)
={\sqrt{3}\kappa^3\hat V^{\frac32}\over10\pi\hat V_{,\Phi}}=
{\sqrt{3}\kappa^3m\phi^2\sqrt{1+(\xx)\psi}\over20\sqrt{2}\pi(1-\psi^2)},
\end{equation}
where all quantities in the RHS 
are calculated at the time $t_k$ when $k$-mode 
leaves the Hubble horizon during inflation, {\it i.e.}, when $k=aH$. 
The large-scale anisotropy of cosmic 
microwave background measured by the COBE-DMR leads \cite{Ben}
\begin{equation}\label{amp}
{\delta T\over T}=\frac{1}{3}\Phi_A(k\simeq a_0H_0)\cong10^{-5},
\end{equation}
where a subscript 0 denotes the present epoch.
The concordant values of $\xi$ and $m$ are also plotted in Fig. 2. 
The spectral index 
$n$ is given by \cite{LL}
\bea\label{index}
n-1&\equiv&{d\ln \Phi_A^2\over d\ln k}
=-{3\hat V_{,\Phi}^{~2}\over\kappa^2\hat V^2}
+{2\hat V_{,\Phi\Phi}\over\kappa^2\hat V}\nn
&=&{4\over\kappa^2\phi^2}\left[-{3(1-\psi)^2\over 1+(\xx)\psi}
+{1-6\psi-(5-35\xi)\psi^2+2(\xx)\psi^3\over\{1+(\xx)\psi\}^2}\right].
\eea
The values of $n$ which satisfy the COBE-DMR normalization are plotted 
in Fig. 3. With the observational data,
$n=1.2\pm0.3$ \cite{Ben}, we have a constraint $|\xi|\lsim10^{-2}$.

To summarize, we have reexamined cosmology of a 
non-minimally coupled massive scalar field. We have shown that inflation 
can occur at the local maximum of the potential in the conformal 
frame, that is, an inflationary universe is created inside a domain
wall just as in the topological inflation scenario of Linde and 
Vilenkin \cite{L-V}. In this case, sufficient inflation is possible even if 
$|\xi|\cong 0.1$, contrary to the previous constraint $|\xi|\lsim 10^{-3}$.
We have also calculated the spectrum of density perturbations. The COBE-DMR 
data requires $|\xi|\lsim 10^{-2}$, which is again much less stringent than that 
claimed by Futamase and Maeda \cite{FM}. In this model our Universe 
exists in a domain which experienced sufficient topological inflation 
and relaxed to $\phi=0$. It is surrounded by a domain wall which 
is eternally inflating with the Hubble parameter 
$H_w=m/(2\sqrt{3|\xi|})$. Beyond the domain wall is an asymptotically de 
Sitter spacetime with a larger expansion rate $H_{as}$.

\acknowledgements

Numerical Computation of this work was 
carried out at the Yukawa Institute Computer Facility. N. S. was supported by 
JSPS Research Fellowships for Young Scientist. This work was supported partially 
by the Grant-in-Aid for Scientific Research Fund of the Ministry of Education, 
Science and Culture No.\ 9702603 (NS), No.\ 09740334 (JY) and 
``Priority Area: Supersymmetry and Unified Theory of Elementary 
Particles (\#707)" (JY).


\baselineskip = 16pt

\begin{figure}
 \begin{center}
  \psbox[height=8.8cm]{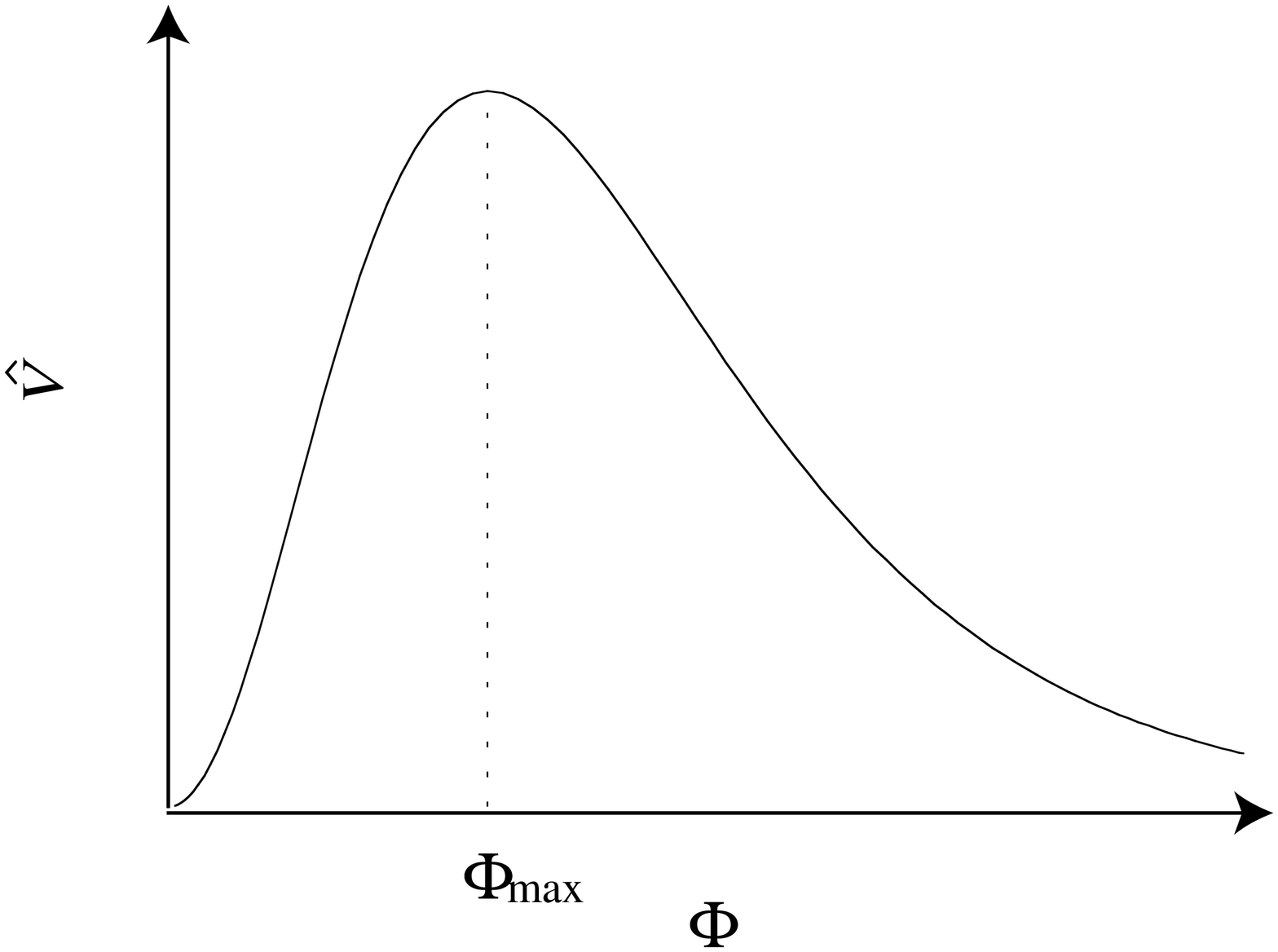}
 \end{center}
\end{figure}
\noindent
{\bf FIG. 1}. The potential $\hat V(\Phi)$ in a fictitious world for 
$\xi<0$ \cite{FM}. A plateau at $\Phi=\Phi_{\rm max}$
leads topological inflation.

\newpage
\begin{figure}
 \begin{center}
  \psbox[height=8.5cm]{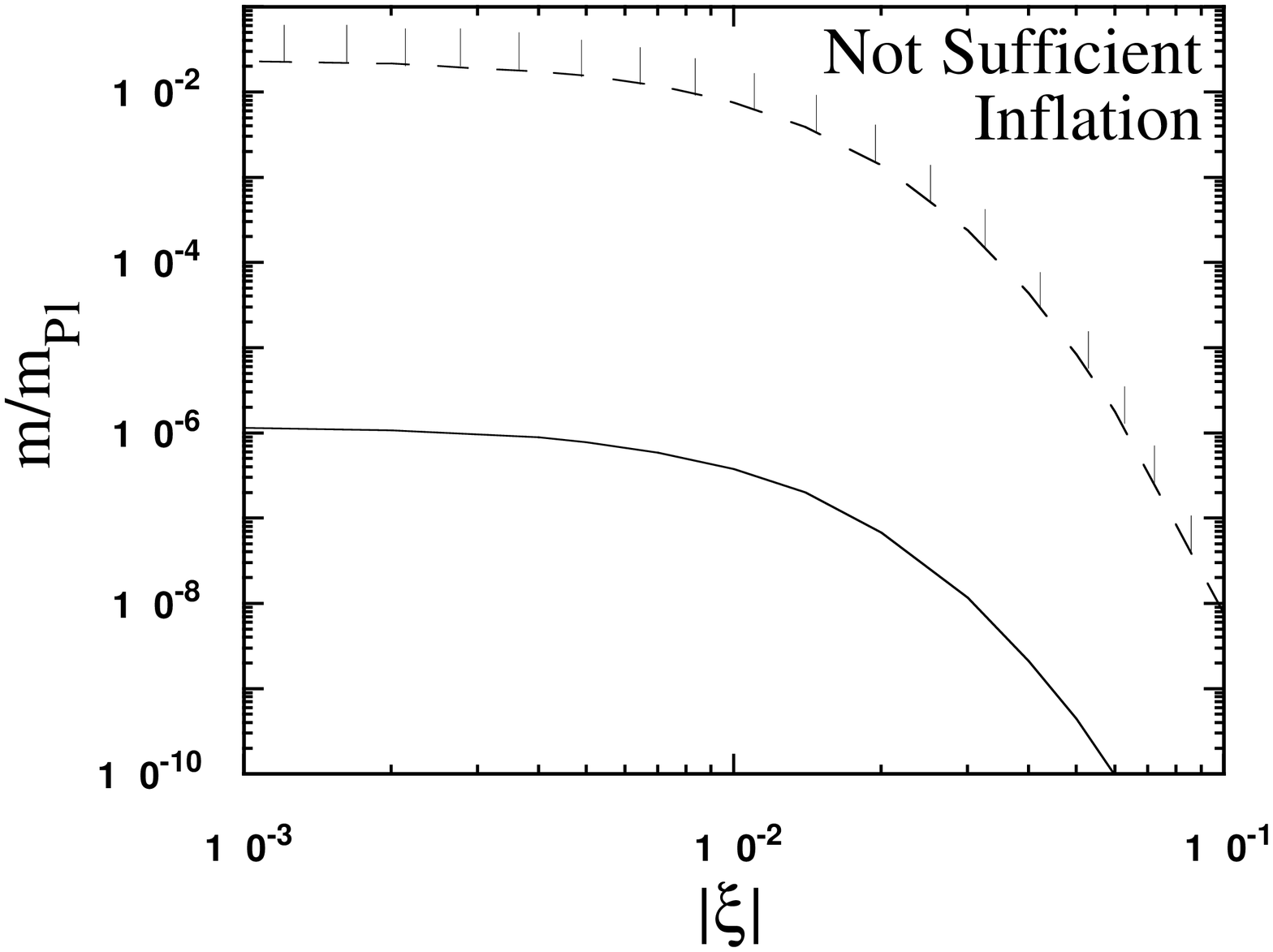}
 \end{center}
\end{figure}

\noindent
{\bf FIG. 2}. Cosmological constraints on $\xi$ and $m/\mpl$. The 
dashed curve is a constraint from sufficient inflation. The 
solid curve represents the concordant values with the 
amplitude of the COBE-DMR data.

\vskip .6cm
\begin{figure}
 \begin{center}
  \psbox[height=8.5cm]{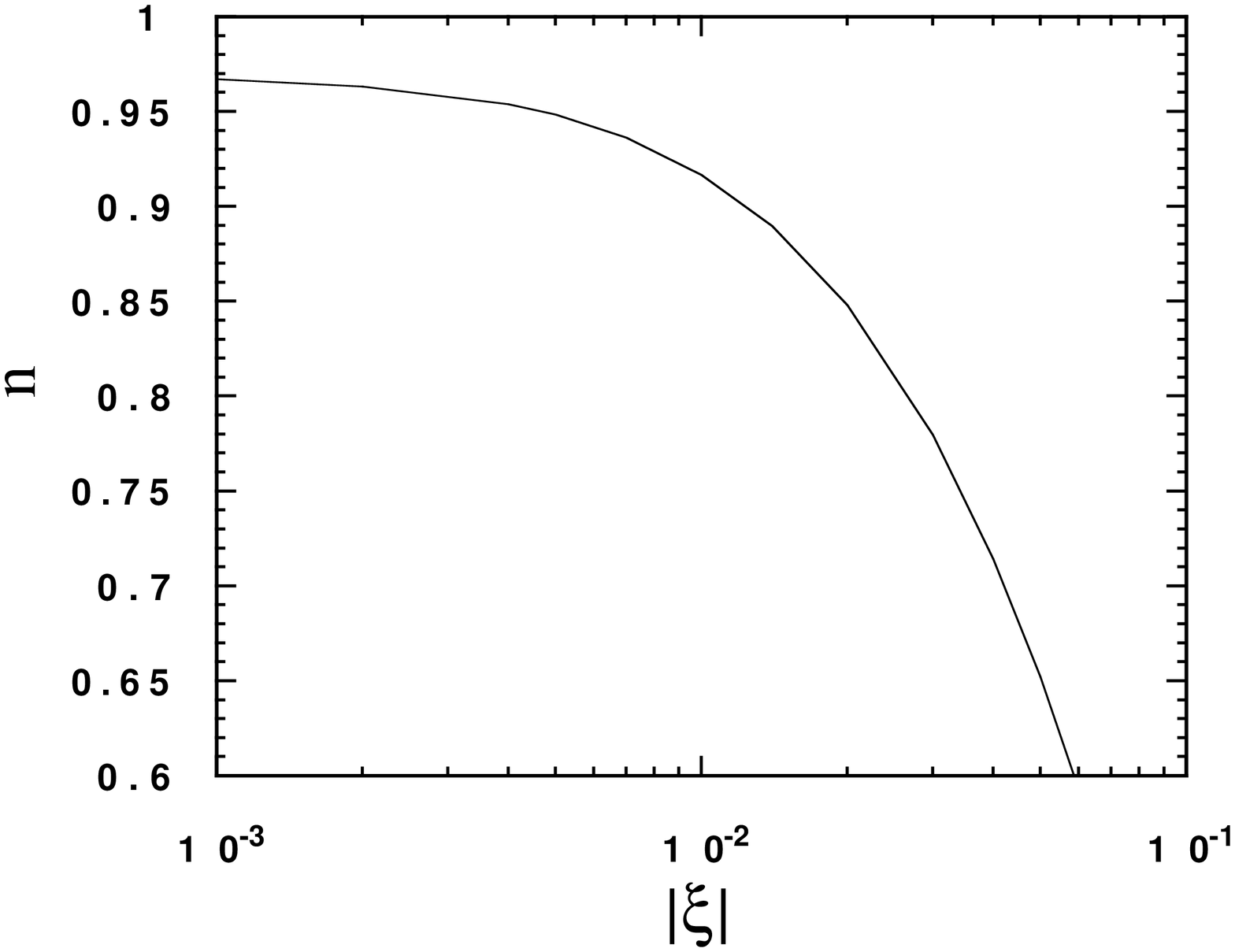}
 \end{center}
\end{figure}

\noindent
{\bf FIG. 3}. The spectral index of density perturbations, which 
satisfies the COBE normalization.

\end{document}